\documentclass{pasj00}
\draft

\begin{document}
\SetRunningHead{K. Muraoka et al.}{NMA Observations of M 83}
\Received{}%{yyyy/mm/dd}
\Accepted{}%{yyyy/mm/dd}

\title{
Nobeyama Millimeter Array Observations of the Nuclear Starburst of M~83: A GMA Scale Correlation between Dense Gas Fraction and Star Formation Efficiency}

%%% Please use the following style in case that sorting by 
%%% affilation is impossible. 
%
 \author{%
   Kazuyuki \textsc{Muraoka},\altaffilmark{1}
   Kotaro \textsc{Kohno},\altaffilmark{2}
   Tomoka \textsc{Tosaki},\altaffilmark{1}
   Nario \textsc{Kuno},\altaffilmark{1}\\
   Kouichiro \textsc{Nakanishi},\altaffilmark{1}
   Toshihiro \textsc{Handa},\altaffilmark{2}
   Kazuo \textsc{Sorai},\altaffilmark{3}
   Sumio \textsc{Ishizuki},\altaffilmark{4}\\
   and Takeshi \textsc{Okuda},\altaffilmark{4}\\
}

 \altaffiltext{1}{Nobeyama Radio Observatory, Minamimaki, Minamisaku, Nagano 384-1305}
 \email{kmuraoka@nro.nao.ac.jp}
 \altaffiltext{2}{Institute of Astronomy, The University of Tokyo, 2-21-1 Osawa, Mitaka, Tokyo 181-0015}
 \altaffiltext{3}{Division of Physics, Graduate School of Science, Hokkaido University, Sapporo 060-0810}
 \altaffiltext{4}{National Astronomical Observatory of Japan, 2-21-1 Osawa, Mitaka, Tokyo 181-8588}
%% `\KeyWords{}' always has to be placed before `\maketitle'.
 %Do NOT move this preamble from here!

\maketitle

\begin{abstract}
We present aperture synthesis high-resolution ($\sim 7^{\prime \prime} \times 3^{\prime \prime}$) observations in
CO($J=1-0$) line, HCN($J=1-0$) line, and 95~GHz continuum emission toward the central ($\sim$ 1.5 kpc) region of
the nearby barred spiral galaxy M~83 with the Nobeyama Millimeter Array.
Our high-resolution CO($J=1-0$) mosaic map depicts the presence of molecular ridges
along the leading sides of the stellar bar and nuclear twin peak structure.
On the other hand, we found the distribution of the HCN($J=1-0$) line emission which traces dense molecular gas
($n_{\rm H_2}$ $>$ a few $\times 10^4$ ${\rm cm}^{-3}$) shows nuclear single peak structure
and coincides well with that of the 95~GHz continuum emission which traces massive starburst.
The peaks of the HCN($J=1-0$) line and the 95~GHz continuum emission are not spatially coincident with 
the optical starburst regions traced by the HST V-band image.
This suggests the existence of deeply buried ongoing starburst due to strong extinction ($A_{\rm V} \sim 5$ mag)
near the peaks of the HCN($J=1-0$) line and the 95~GHz continuum emission.
We found that the HCN($J=1-0$)/CO($J=1-0$) intensity ratio $R_{\rm HCN/CO}$ correlates well
with extinction-corrected SFE in the central region of M~83 at a resolution of $7^{\prime \prime}.5$ ($\sim 160$ pc).
This suggests that SFE is controlled by dense gas fraction traced by $R_{\rm HCN/CO}$
even on a Giant Molecular cloud Association (GMA) scale.
Moreover, the correlation between $R_{\rm HCN/CO}$ and the SFE in the central region of M~83 seems to be
almost coincident with that of the \citet{gao2004a} sample.
This suggests that the correlation between $R_{\rm HCN/CO}$ and the SFE on a GMA ($\sim$ 160 pc) scale
found in M~83 is the origin of the global correlation on a few kpc scale shown by \citet{gao2004a}.
\end{abstract}

\section{Introduction}

Star formation is one of the most fundamental processes of the evolution of galaxies.
Stars are formed by the contraction of molecular interstellar medium (ISM),
and then a new ISM containing heavy element is dispersed into interstellar space
by stellar wind and supernova explosions when stars end their lives.
By repeating these processes, ISM, galaxies, and the Universe evolve.
Therefore, the physics of star formation is very important in the understanding of
the evolution of galaxies.

In the central regions of disk galaxies, we often find intense star-formation activities
(e.g., \cite{ho1997}) where not only the star formation rate (SFR)
but also the star formation efficiency (SFE; \cite{young1996}), which is an intensive parameter defined as
the fraction of the SFR surface density in the surface mass density of molecular gas, are also enhanced.
These star formation activities in central regions of galaxies are different from a simply scaled-up version
of that in the disk region, and it is are referred to as starburst \citep{kennicutt1998b}.
The nuclear starburst is prominent not only because they show high SFR but also because they show elevated SFE.
Therefore, for understanding the nuclear starburst we should reveal what causes the high SFE star formation.

It is essential to study {\it dense} molecular gas,
because stars are formed from the dense cores of molecular clouds.
In fact, recent studies of the dense molecular medium in galaxies based on the observations in the HCN($J=1-0$) emission,
a dense molecular gas tracer ($n_{\rm H_2}$ $>$ a few $\times 10^4$ ${\rm cm}^{-3}$) due to
its large dipole moment ($\mu_{\rm HCN} = 3.0$ Debye, whereas $\mu_{\rm CO} = 0.1$ Debye),
demonstrate the intimate connection between dense molecular gas and massive star formation in galaxies.
After the pioneer work by \citet{solomon1992}, a tight correlation between the HCN($J=1-0$) intensity and 
FIR continuum luminosities has been shown among many galaxies including nearby ones
(\cite{gao2004a}, \cite{gao2004b}), and high redshift quasar host galaxies (\cite{carilli2005}, \cite{wu2005}),
although some deviations from the correlation are reported toward extreme environments (\cite{riechers2007}; \cite{gracia2008}, \cite{kohno2008a}).
A spatial coincidence between dense molecular gas traced by the HCN($J=1-0$) emission and
massive star-forming regions was also reported (\cite{kohno1999}, \cite{shibatsuka2003}).
In addition, \citet{gao2004a} reported that HCN($J=1-0$)/CO($J=1-0$) integrated intensity ratio in the brightness temperature scale,
hereafter $R_{\rm HCN/CO}$, correlates well with SFE.
This suggests that SFE is controlled by the fraction of dense molecular gas
to the total amount of molecular contents traced by $R_{\rm HCN/CO}$
(e.g., \cite{solomon1992}, \cite{kohno2002}, \cite{shibatsuka2003}, \cite{gao2004a}).

However, these previous reports on the relationship between dense gas fraction and SFE are derived
from coarse spatial resolution ($> 30^{\prime \prime}$) data of moderately distant galaxies ($D > 10$ Mpc).
It means the results show a correlation on global scales of galaxies.
In our Galaxy, a massive star-forming region is associated with the giant molecular clouds (GMCs)
or their associations (giant molecular associations; GMAs).
A typical scale of a GMC is a few 10 pc (\cite{scoville1987}) and that of a GMA is a few 100 pc. Therefore, we should 
investigate a correlation between dense gas fraction and SFE on this scale.
To address this issue, observations with high-angular resolution of the central region of nearby galaxies are required.

In this paper, we present CO($J=1-0$), HCN($J=1-0$), and 95~GHz continuum observations
toward the central regions of M~83 (NGC~5236) using the Nobeyama Millimeter Array (NMA).
M~83 is a nearby, face-on, barred, grand-design spiral galaxy hosting an intense starburst at its center.
The distance to M~83 is estimated to be 4.5 Mpc \citep{thim2003};
therefore, 1$^{\prime \prime}$ corresponds to 22 pc.
We can then discuss the relation between $R_{\rm HCN/CO}$ and the SFE at the center of M~83 on a GMA scale
with a few arcsec resolution observations, which can be accomplished with the NMA.
The high-resolution ($\leq$ 23$^{\prime \prime}$ $\sim$ 500 pc) CO line observations of the central region
of M~83 have been performed repeatedly with various telescopes and in various transitions.
We summarized them in table~1.
For example, \citet{handa1990} mapped the distribution of CO($J=1-0$) emission for the center and bar
at a resolution of $16^{\prime \prime}$ using the NRO 45-m telescope.
They observed a large concentration of molecular gas in the nucleus and an extended ridge along the major axis of the bar.
\citet{ishizuki1993} reported the CO($J=1-0$) observation toward the center with the NMA.
The distribution of CO($J=1-0$) emission shows twin peak structure at the nucleus.
\citet{sakamoto2004} performed CO($J=3-2$) and CO($J=2-1$) emission observations toward the center and the bar
using the Submillimeter Array (SMA). Their CO maps revealed two gas ridges on the leading side
of the stellar bar and a nuclear gas ring of $\sim$ 300 pc in diameter.
The distribution of CO($J=3-2$) emission is not coincident with the nuclear starburst,
which is depicted by a three-color composite image made from $HST$/WFPC2 data,
F300W, F547M, and F814W.
The authors mentioned that the nuclear starburst in M~83 is lopsided, mostly
on the receding side (south) of the dynamical center.

\begin{table}
\begin{center}
Table~1.\hspace{4pt}Previous high-resolution CO observations of M~83.\\[1mm]
\begin{tabular}{crlll}
\hline \hline \\[-3mm]
Transition & Authors\,\,\,\,\,\,\,\,\,\,\,\,\,\,\,\,\,\,\,\,\,\, &\,\,\, Telescope & map size & resolution \\[1mm]
\hline \\[-3mm]
J=1--0 & \citet{handa1990} & NRO 45-m & 3$^{\prime}$.5 $\times$ 1$^{\prime}$ &16$^{\prime \prime}$  \\[1.0mm]
       & \citet{ishizuki1993} & NMA & 1$^{\prime}$.1 $\times$ 1$^{\prime}$.1 &12$^{\prime \prime}$ $\times$ 6$^{\prime \prime}$ \\[1.0mm]
       & \citet{handa1994} & NMA & 1$^{\prime}$.1 $\times$ 1$^{\prime}$.1 &12$^{\prime \prime}$ $\times$ 6$^{\prime \prime}$ \\[1.0mm]
       & \citet{kuno2007} & NRO 45-m & 6$^{\prime}$ $\times$ 6$^{\prime}$ &16$^{\prime \prime}$ \\[1.0mm]
       & This work        & NMA  & 2$^{\prime}$ $\times$ 1$^{\prime}$ & 7$^{\prime \prime}$ $\times$ 3$^{\prime \prime}$ \\[1.0mm]
\hline \\[-3mm]
J=2--1 & \citet{wall1991} & JCMT 15 m &  &22$^{\prime \prime}$  \\[1.0mm]
       & \citet{israel2001} & JCMT 15 m & 1$^{\prime}$.2 $\times$ 2$^{\prime}$ &21$^{\prime \prime}$  \\[1.0mm]
       & \citet{lundgren2004} & SEST 15 m & 10$^{\prime}$ $\times$ 10$^{\prime}$ &23$^{\prime \prime}$ \\[1.0mm]
       & \citet{sakamoto2004} & SMA & 1$^{\prime}$.5 $\times$ 1$^{\prime}$ &3$^{\prime \prime}$.8 $\times$ 2$^{\prime \prime}$.5  \\[1.0mm]
\hline \\[-3mm]
J=3--2 & \citet{wall1991} & CSO 10 m &  &22$^{\prime \prime}$  \\[1.0mm]
       & \citet{petitpas1998} & JCMT 15 m & 0$^{\prime}$.7 $\times$ 0$^{\prime}$.7 &14$^{\prime \prime}$  \\[1.0mm]
       & \citet{israel2001} & JCMT 15 m & 1$^{\prime}$.2 $\times$ 1$^{\prime}$.7 &14$^{\prime \prime}$  \\[1.0mm]
       & \citet{dumke2001} & SMTO 10 m & 2$^{\prime}$ $\times$ 2$^{\prime}$ &22$^{\prime \prime}$  \\[1.0mm]
       & \citet{sakamoto2004} & SMA & 0$^{\prime}$.6 $\times$ 0$^{\prime}$6 &3$^{\prime \prime}$.1 $\times$ 1$^{\prime \prime}$.5 \\[1.0mm]
       & \citet{bayet2006} & CSO 10 m & 4$^{\prime}$.5 $\times$ 2$^{\prime}$ &22$^{\prime \prime}$  \\[1.0mm]
       & \citet{muraoka2007} & ASTE 10 m & 5$^{\prime}$ $\times$ 5$^{\prime}$ &22$^{\prime \prime}$  \\[1.0mm]
\hline \\[-3mm]
J=4--3 & \citet{petitpas1998} & JCMT 15 m & 0$^{\prime}$.6 $\times$ 0$^{\prime}$.5 &11$^{\prime \prime}$  \\[1.0mm]
       & \citet{israel2001} & JCMT 15 m & 0$^{\prime}$.5 $\times$ 0$^{\prime}$.5 &11$^{\prime \prime}$  \\
\hline \\[-2mm]
\end{tabular}\\
{\footnotesize
}
\end{center}
\end{table}

It is difficult to estimate the star formation activity accurately in the central starburst region because of strong extinction.
\citet{harris2001} presented a photometric catalog of 45 massive star clusters
in the nuclear starburst region of M~83, observed with $HST$/WFPC2.
They revealed that the ages of star clusters in the central 300~pc are younger than 10~Myr old,
and the cluster age distribution has an extremely sharp peak at 5 -- 7 Myr.
However, the cluster ages in the north side of the nucleus are unknown
since no clusters were found due to strong extinction by dust.
They also mentioned the original starburst had started at the southern end of
the nuclear starburst region, and propagated northward.
The propagation of star formation is also reported by \citet{ryder2005} and by \citet{houghton2008},
who quantify the age gradient better using near-IR data.

To the north side of the nucleus, the H$\alpha$/H$\beta$ ratio is higher than that to the south side
(see figure~4 in \cite{harris2001}). This means that dust extinction to the north side is stronger than that to the south side.
The strong dust extinction indicates the existence of huge amounts of dust and molecular gas.
In addition, the extinction on the north side has not yet been accurately estimated since it is too strong,
and several authors suggests the existence of a highly-obscured nuclear starburst (\cite{ryder2005}, \cite{fathi2008}, \cite{houghton2008}).
Therefore, it is necessary to evaluate the strength of extinction to investigate
the relationship between molecular gas and true star formation in the central region of M~83.

The goals of this paper are: (1) to obtain distributions of total molecular gas and dense molecular gas through 
CO($J=1-0$) line and HCN($J=1-0$) line emission in the central 1.5 kpc region of M~83 with a spatial resolution of $\sim$ 100 pc,
(2) to evaluate the extinction and reveal its distribution, and to obtain the extinction-corrected SFR and SFE.
(3) to examine the correlation between dense gas fraction traced by $R_{\rm HCN/CO}$ and the SFE on a GMA scale.

\section{Observations and Data Reduction}

Aperture synthesis observations in the CO($J=1-0$) line, the HCN($J=1-0$) line, and the 95~GHz continuum emission 
towards the central region ($\sim$ 1.5 kpc) of M~83 were carried out with the NMA
during the periods from December 2003 to April 2004, and from December 2005 to April 2006.
CO($J=1-0$) line observations were made toward 3 adjacent field-of-views (FOVs) to make a mosaic of the center of the galaxy.
HCN($J=1-0$) line and 95~GHz continuum observations were made toward only a central FOV.
Figure~1 shows the each FOV superposed on the V-band image obtained with the VLT \citep{comeron2001}.

The NMA consists of six 10-m antennas equipped with cryogenically cooled receivers employing
Superconductor-Insulator-Superconductor mixers in double side band operation.
Three antenna configurations (AB, C, and D) were used during the observations.
The backend used was the Ultra Wide-Band Correlator \citep{okumura2000}.

A radio source, J1337-129, was observed every $\sim$ 20 minutes for amplitude
and phase calibrations, and the passband shape of the system was determined from observations of a strong
continuum source 3C 273. The flux density of J1337-129 was measured at several times during an observing run
based on that of a quasar 3C 345. Its flux is determined by flux measurements of Saturn and Uranus.
The overall error of the absolute flux scale was estimated to about $\pm$20\%.

The raw visibilities were edited and calibrated using the NRO UVPROC-II package \citep{tsutsumi1997},
and then final images were created using the IMAGR task in the NRAO AIPS package.
For some spectral line analysis, data also have the continuum emission.
In this case, we subtracted the continuum emission from the visibilities using the UVPROC-II task LCONT at first.
We use the subtracted continuum emission in HCN($J=1-0$) data as the 95~GHz continuum emission.
Therefore, this flux is actually an average flux in two separated bands at 88.741 $\pm$ 0.5~GHz and 100.741 $\pm$ 0.5~GHz.
Mosaic images in the CO($J=1-0$) line were made using MIRIAD \citep{sault1995}.
Achieved sensitivities and spatial resolutions are summarized in table~2 and 3.

%% fig. 1
\begin{figure}
  \begin{center}
    \FigureFile(80mm,75mm){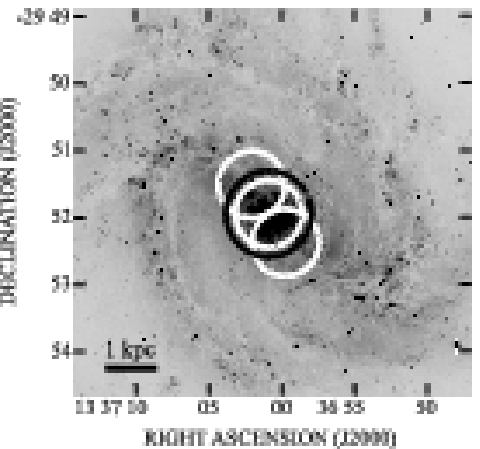}
  \end{center}
\caption{
Observed FOVs with the NMA superposed on a VLT V-band image of M~83 \citep{comeron2001}.
White circles indicate the FOV for CO($J=1-0$) observations,
and a black circle indicates that for HCN($J=1-0$) observations.
}
\label{fig:fig1}
\end{figure}
%% fig. 1

%%%%%%%%%%%
% Table~2

\begin{table}
\begin{center}
Table~2.\hspace{4pt}Observation parameters and results of M~83 with the NMA.\\[1mm]
\begin{tabular}{llll}
\hline \hline \\[-3mm]
Observed line & & CO($J=1-0$) & HCN($J=1-0$)\\[1mm]
\hline \\[-3mm]
Rest frequency & [GHz] & 115.271204 & 88.631604 \\
Band width & [MHz] & 512 & 1024\\
Band resolution & [MHz] & 2 & 8\\
Velocity resolution & [km s$^{-1}$] & 5.2 & 27.1\\
Sideband & & upper & lower\\
FOV & [$^{\prime \prime}$]& 59 & 77\\
Number of FOVs & & 3 (mosaic) & 1\\
Observed date&  & 2003/12-2004/4 & 2005/12-2006/4\\
Synthesized beam size & [$^{\prime \prime}$] & 7.2 $\times$ 3.4 & 7.1 $\times$ 3.1\\
Equivalent $T_{\rm b}$  & [K (Jy\,beam$^{-1}$)$^{-1}$] & 3.76 & 7.15 \\
rms noise (channel map) & [mJy\,beam$^{-1}$] & 85 & 7.2  \\
                        & [mK] & 320 & 51.5 \\
rms noise (intensity map) &[Jy\,beam$^{-1}$\,km\,s$^{-1}$] & 2.7 & 0.52\\
                          & [K km s$^{-1}$] & 10.2 & 3.72 \\
total flux within FOV   & [Jy\,km\,s$^{-1}$] & 2.4$\pm$0.1 $\times$ 10$^3$ & 60$\pm$10 \\
\hline \\[-4mm]
\multicolumn{2}{l}{The central position of the FOV$^{\ast}$} & & \\
\,\,\,\,\,\,R.A. (J2000) & \multicolumn{2}{l}{$13^{\rm h} 37^{\rm m} 00^{\rm s}.90$} & \\
\,\,\,\,\,\,Decl. (J2000) & \multicolumn{2}{l}{$-29^{\circ} 51^{\prime} 56^{\prime \prime}.7$} & \\
\hline \\[-2mm]
\end{tabular}\\
{\footnotesize
$^{\ast}$ Reference of the central position of the FOV --- \citet{jarrett2003}
}
\end{center}
\end{table}

%%%%%%%%%%%
% Table~3

\begin{table}
\begin{center}
Table~3.\hspace{4pt}Summary of the 95~GHz continuum in M~83.\\[1mm]
\begin{tabular}{llc}
\hline \hline \\[-3mm]
Synthesized beam & [$^{\prime \prime}$] & 8.0 $\times$ 3.4 \\
Equivalent $T_{\rm b}$ & [K (Jy\,beam$^{-1}$)$^{-1}$] & 5.00 \\
rms noise & [mJy\,beam$^{-1}$] &  0.85 \\
          & [mK]               &  4.25 \\
total flux within FOV & [mJy]  &  30   \\
\hline \\[-2mm]
\end{tabular}\\
{\footnotesize Here, 95 GHz means a combination of 88.741$\pm$0.5 GHz and 100.741$\pm$0.5 GHz.}
\end{center}
\end{table}

\section{Results}

\subsection{Maps}

Figure~2 shows an integrated intensity map and a velocity field in the CO($J=1-0$) line,
an integrated intensity map in the HCN($J=1-0$) line, and a map of the 95~GHz continuum emission.
Our high resolution CO($J=1-0$) mosaic map depicts the presence of molecular ridges
along the leading sides of the stellar bar and nuclear twin-peaks structure.
This structure is seen in the previous studies in the CO($J=1-0$) (\cite{handa1994}), and
in the CO($J=2-1$) and the CO($J=3-2$) \citep{sakamoto2004}.
Strong non-circular motion along the molecular ridges was also detected in the velocity field.
This strong non-circular motion is consistent well to that detected in the CO($J=2-1$) velocity field (\cite{sakamoto2004}).
The motion might play an important role in feeding large amount of molecular gas into the starburst nucleus.
We determined the dynamical center from the CO velocity field using the AIPS task GAL (see table~4).
We find the distributions of the HCN($J=1-0$) line intensity,
and the 95~GHz continuum flux are confined to a single peak, corresponding to the northern peak seen in the CO map.
This means that dense molecular gas traced by the HCN($J=1-0$) line and current star formation traced
by the 95~GHz continuum emission are distributed only at north side of the center,
although less dense molecular gas seen in the CO($J=1-0$) line emission are widely distributed in the central region.
The peak position of HCN($J=1-0$) line intensity is consistent spatially with the youngest star clusters
found by \citet{harris2001} and \citet{houghton2008}.

%% fig. 2
\begin{figure}
  \begin{center}
    \FigureFile(170mm,181mm){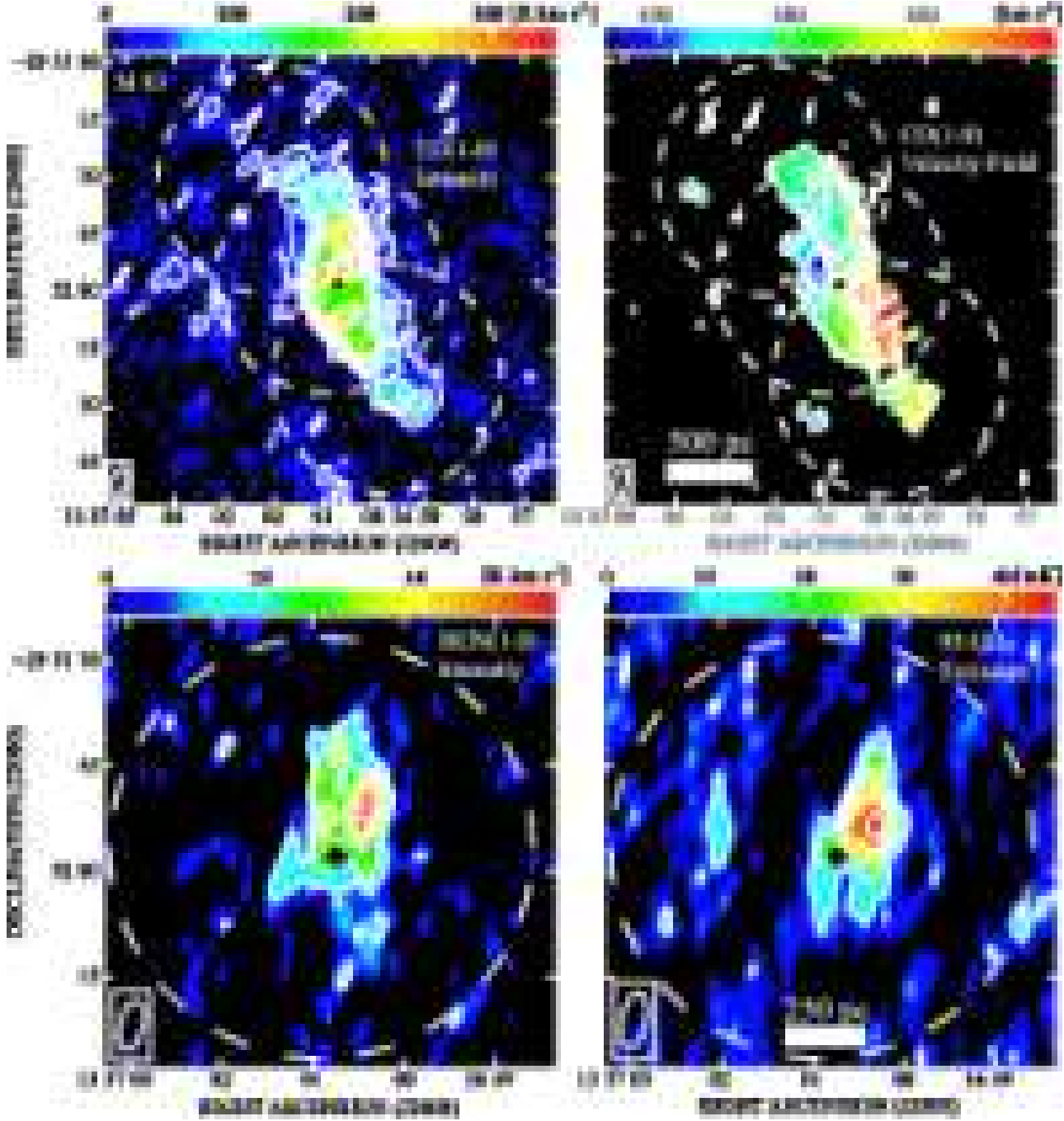}
  \end{center}
\caption{
(top left) Integrated intensity map in the CO($J=1-0$) emission in the central region of M~83.
The central cross indicates the dynamical center determined from our CO velocity field,
and dashed white circles indicate the FOVs of CO($J=1-0$) observations.
The synthesized beam is shown in the lower left corner.
The contour levels are 3, 6, 9, 12, 15, 21, 27, and 33 $\sigma$,
where 1 $\sigma$ = 2.7 Jy\,beam$^{-1}$\,km\,s$^{-1}$ or 10 K km s$^{-1}$.
(top right) Velocity field measured in CO($J=1-0$). The contour levels are from 440 to 590 km s$^{-1}$
with an interval of 10 km s$^{-1}$.
(bottom left) Integrated intensity map in the HCN($J=1-0$) emission in the central region of M~83.
The contour levels are 3, 6, 9, 12, and 15 $\sigma$,
where 1 $\sigma$ = 0.52 Jy\,beam$^{-1}$\,km\,s$^{-1}$ or 3.7 K km s$^{-1}$.
(bottom right) Map of the 95~GHz continuum emission in the central region of M~83.
The contour levels are 2, 4, 6, 8, 10 $\sigma$, where 1 $\sigma$ = 4.3 mK or 0.85 mJy\,beam$^{-1}$.
}
\label{fig:fig2}
\end{figure}
%% fig. 2

Here, we summarized the positions of several peaks and nuclei
determined from near-IR and our mm-wave data in figure~3 and table~4.
The dymanical center determined from our CO velocity field is near
the symmetry center of the outer K-band isophotes,
whereas the visible nucleus (that is also the location of the K-band photometric peak in table~4) is offset from the dynamical center and the peaks of HCN($J=1-0$) line intensity, 95~GHz continuum emission,
and the extinction-corrected H$\alpha$ luminosity (see section 4.1).
The position of the hidden mass concentration discussed in \citet{diaz2006} and \citet{houghton2008} seems
to be coincident with the peaks of HCN($J=1-0$) line intensity and the 95~GHz continuum emission,
although the spatial resolutions of our mm-wave data ($7^{\prime \prime} \times 3^{\prime \prime}$) are insufficient
to make a detailed comparison.

We compare our HCN($J=1-0$) map with that obtained by \citet{helfer1997}.
The peak positions of the HCN($J=1-0$) line emission are almost coincident between these two maps,
whereas the flux values of the peak position are different.
When our map was convolved to $12^{\prime \prime}.5 \times 4^{\prime \prime}.1$, which is the beam size of their map,
the peak flux was measured as 12.5 Jy\,km\,s$^{-1}$. This value is about 30\% lower than that of their data.
It is unclear what causes such a significant discrepancy.
Note that our HCN($J=1-0$) line flux measured with the NMA is consistent well with that measured with the NRO 45-m telescope
for central 22$^{\prime \prime}$ region (see section 3.2).

%% fig. 3
\begin{figure}
  \begin{center}
    \FigureFile(80mm,81mm){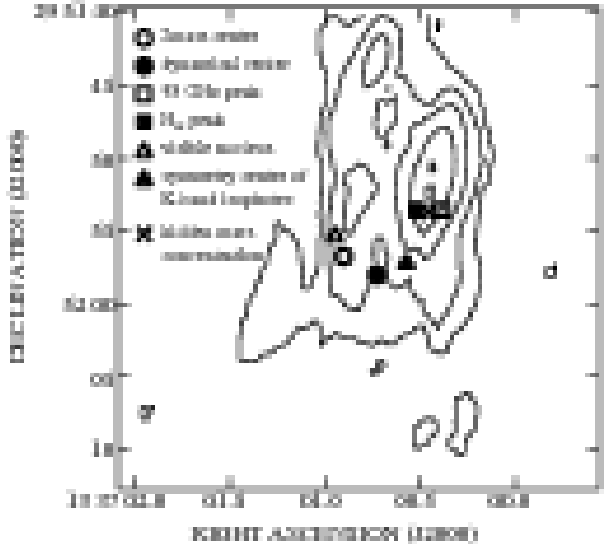}
  \end{center}
\caption{Several peaks/nuclei in the central region of M~83 superposed on the HCN($J=1-0$) integrated intensity map.
The contour levels are the same as those in the bottom left of figure~2.
The open circle shows infrared peak identified by the 2MASS image (the central position of the FOV),
the filled circle shows the dynamical center,
the open square shows the 95 GHz continuum peak,
the filled square shows the peak of extinction-corrected H$\alpha$ luminosity (see section 4.1),
the open triangle shows the visible nucleus,
the filled triangle shows the symmetry center of the outer K-band isophotes,
and the x mark shows the hidden mass concentration.
See table~4 and text for details.
}
\label{fig:fig3}
\end{figure}
%% fig. 3

\begin{table}
\begin{center}
Table~4.\hspace{4pt}Peaks/nuclei in the central region of M~83\\[1mm]
\begin{tabular}{llll}
\hline \hline \\[-5mm]
R.A. & Decl. & description & reference \\
\hline \\[-5mm]
13 37 00.90 & -29 51 56.7 & 2MASS, (the central position of the FOV)  & \citet{jarrett2003} \\
13 37 00.73 & -29 51 57.9 & dynamical center determined from CO velocity field & this work \\
13 37 00.38 & -29 51 53.5 & 95 GHz continuum peak & this work \\
13 37 00.52 & -29 51 53.5 & peak of extinction-corrected H$\alpha$ luminosity & this work \\
\hline \\[-5mm]
13 37 00.95 & -29 51 55.5 & K-band photometric peak, visible nucleus & \citet{thatte2000} \\
13 37 00.57 & -29 51 56.9 & the symmetry center of the outer K-band isophotes & \citet{thatte2000} \\
13 37 00.46 & -29 51 53.6 & hidden mass concentration & \citet{diaz2006} \\
\hline \\[-5mm]
\end{tabular}\\
{\footnotesize
}
\end{center}
\end{table}

\subsection{Combining CO($J=1-0$) data obtained with the NMA and the NRO 45-m}

We found that the CO($J=1-0$) spectrum at the center of M~83 obtained with the NMA is
different from that obtained with the NRO 45-m telescope \citep{kuno2007}
at the same resolution of 16$^{\prime \prime}$ as shown in figure~4.
This corresponds to missing flux of the interferometry,
and we evaluated that the missing flux was $\sim$ 30 \%.
In order to correct the missing flux and obtain the true flux value in the CO($J=1-0$) emission with the NMA,
we combined the NMA CO($J=1-0$) data with the NRO 45-m CO($J=1-0$) data.

%% fig. 4
\begin{figure}
  \begin{center}
    \FigureFile(80mm,87mm){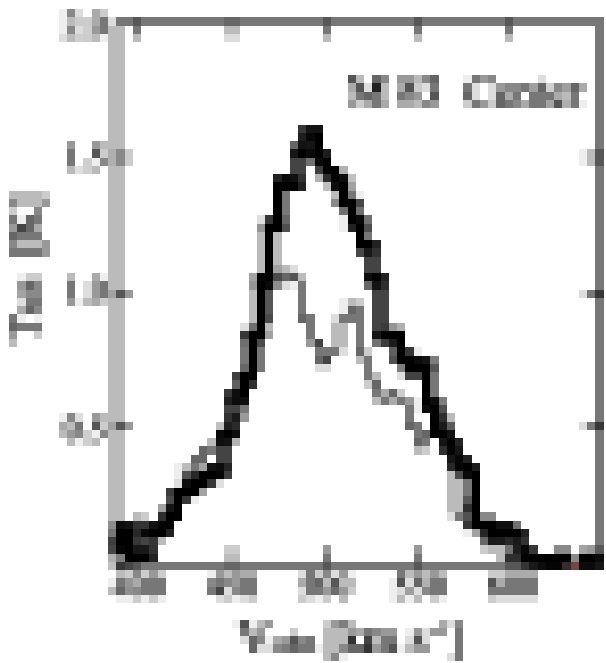}
  \end{center}
\caption{
CO($J=1-0$) spectra at the center of M~83. The thick line corresponds to the spectrum
obtained with the NRO 45-m \citep{kuno2007}, and the thin line corresponds to that with the NMA (this work),
which is convolved to the resolution of 16$^{\prime \prime}$ to match the NRO 45-m data.
The velocity resolutions of these spectra are both 5.2 km\,s$^{-1}$.
The spectrum obtained with the NMA does not have a single peak but a twin-peak profile,
and its intensity is $\sim$ 30\% lower than that obtained with the NRO 45-m.
}
\label{fig:fig4}
\end{figure}
%% fig. 4

We employed the latest CO($J=1-0$) data (Fukuhara et al. in prep.) obtained with the NRO 45-m
using the On-the-Fly (OTF) method.
The combination of two CO($J=1-0$) data sets was performed using MIRIAD.
Figure~5 shows a combined CO($J=1-0$) image and an CO($J=1-0$) image only with the NMA data.
The results and specification of the combining are summarized in table~5.
Compared to the NMA-only image, the flux was properly recovered
at the center, and the CO-emitting area is enlarged in the combined image.
That is, emission from diffuse components of molecular gas could be reproduced appropriately.
The 45-m data made the resultant noise level lower and the synthesized beam size slightly wider.
However, the overall structure at the center,
such as the twin-peaks (or ring-like) is little affected by this combining process.
For the HCN ($J=1-0$) line, we did not combine the NMA data with single dish data.
It is because we could not find any missing flux in the HCN($J=1-0$) flux
for the central 22$^{\prime \prime}$ region. The flux obtained with the NMA, 38 $\pm$ 3 Jy\,\,km\,s$^{-1}$,
coincides well with that obtained with the NRO 45-m, 39 $\pm$ 2 Jy\,\,km\,s$^{-1}$ (Hirota et al. in prep).
This means that there is no significant diffuse component of HCN($J=1-0$) emission.
The difference in missing flux between CO($J=1-0$) flux and HCN($J=1-0$) flux
suggests HCN emitting region is more restricted than CO emitting region.
This is because HCN($J=1-0$) emission is originated from dense molecular gas ($n_{\rm H_2}$ $>$ a few $\times 10^4$ ${\rm cm}^{-3}$)
associating with star-forming region, whereas CO($J=1-0$) emission is originated from low-dense ($n_{\rm H_2}$ $\sim$ $\times 10^2$ ${\rm cm}^{-3}$)
component of molecular gas which is distributed ubiquitously.

%% fig. 5
\begin{figure}
  \begin{center}
    \FigureFile(170mm,93mm){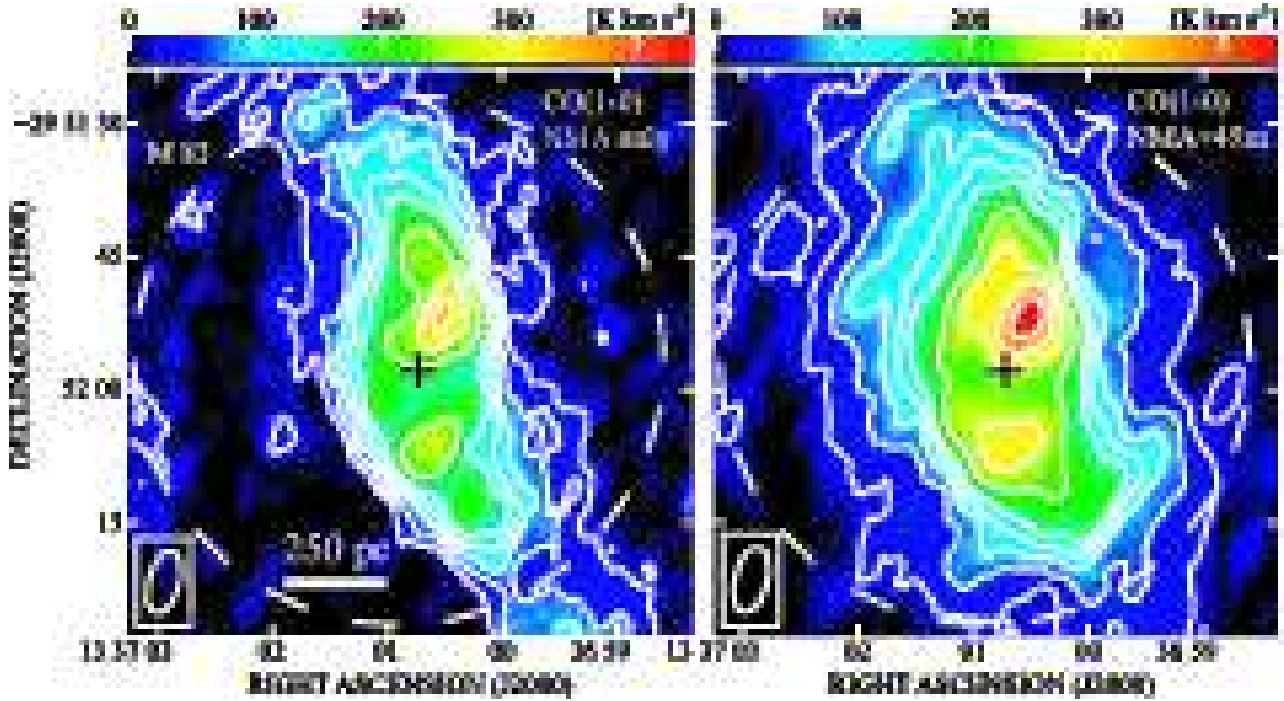}
  \end{center}
\caption{
(left) Integrated intensity map in the CO($J=1-0$) before combining.
The contour levels and the peak are the same as those in the top left of figure~2.
The central cross indicates the dynamical center (see table~4).
(right) Integrated intensity map in the CO($J=1-0$) after combining.
The contour levels are 3, 6, 9, 12, 15, 21, 27, 33, and 39 $\sigma$,
where 1 $\sigma$ = 2.7 Jy\,beam$^{-1}$\,km\,s$^{-1}$ or 10 K km s$^{-1}$.
}
\label{fig:fig5}
\end{figure}
%% fig. 5

\begin{table}
\begin{center}
Table~5.\hspace{4pt}Parameters of CO($J=1-0$) images before/after the 45-m data combine\\[1mm]
\begin{tabular}{llcc}
\hline \hline \\[-3mm]
 & & NMA only & NMA + 45-m \\[1mm]
\hline \\[-3mm]
rms noise (channel map)  & [mK] & 320 & 273 \\
                         & [mJy\,beam$^{-1}$] & 85 & 80 \\
rms noise (intensity map)& [K km s$^{-1}$] & 10.2 & 8.53 \\
                         & [Jy\,beam$^{-1}$ km s$^{-1}$] & 2.7 & 2.5 \\
Synthesized beam size    & [$^{\prime \prime}$] & 7.2 $\times$ 3.4 & 7.5 $\times$ 3.6 \\
peak flux                & [K km s$^{-1}$] & 350 & 442 \\
                         & [Jy\,beam$^{-1}$ km s$^{-1}$] & 93 & 120 \\
Equivalent $T_{\rm b}$  & [K (Jy\,beam$^{-1}$)$^{-1}$] & 3.76 & 3.41 \\
\hline \\[-2mm]
\end{tabular}\\
{\footnotesize
}
\end{center}
\end{table}

We compare the combined CO($J=1-0$) image with the NMA HCN($J=1-0$) image in figure~6.
This corresponds to the comparison between the total amount of the molecular gas and
the dense molecular gas which will host star formation activity soon.
The peak of the HCN($J=1-0$) intensity is almost coincident with the northern peak of CO($J=1-0$) intensity,
whereas little HCN($J=1-0$) emission was seen at the southern peak of CO($J=1-0$).
Smaller fraction of the dense gas at the southern peak suggests that this region may be in a post star formation phase.

%% fig. 6
\begin{figure}
  \begin{center}
    \FigureFile(80mm,81mm){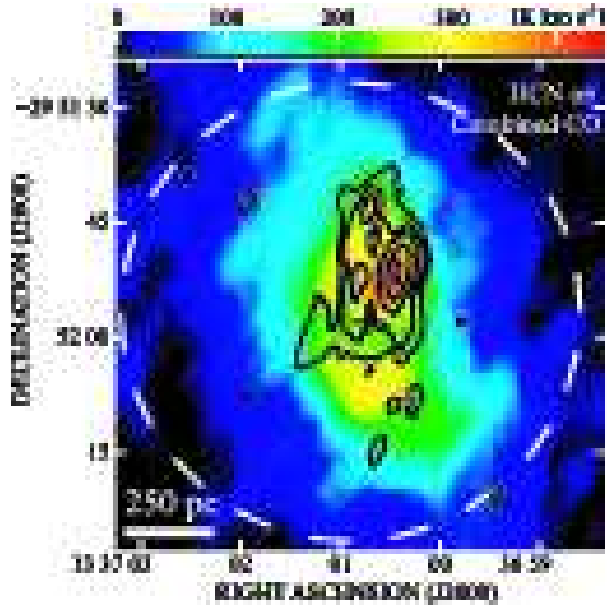}
  \end{center}
\caption{Integrated intensity map in the HCN($J=1-0$) emission (contour) superposed on
an integrated intensity map of the combined CO($J=1-0$) data (color).
The central cross indicates the dynamical center.
The contour levels of HCN($J=1-0$) emission are the same as those of figure~2.
The peak of HCN($J=1-0$) intensity is almost coincident with the northern peak of CO($J=1-0$) intensity,
whereas little HCN($J=1-0$) emission was seen at the southern peak of CO($J=1-0$).
}
\label{fig:fig6}
\end{figure}
%% fig. 6

\subsection{HCN($J=1-0$)/CO($J=1-0$) intensity ratio $R_{\rm HCN/CO}$}

We made an intensity ratio map from the HCN($J=1-0$) data and the combined CO($J=1-0$) data.
To improve the data quality and to adjust the beamsize we convolved the images to have
the same $7^{\prime \prime}.5 \times 7^{\prime \prime}.5$ (160 pc $\times$ 160 pc) resolution
before calculating the ratio. The resultant $R_{\rm HCN/CO}$ map is shown in figure~7.
The peak of $R_{\rm HCN/CO}$ value is 0.11 $\pm$ 0.01.
The region with high $R_{\rm HCN/CO}$ is concentrated to the north side of the dynamical center of the galaxy.
The high ratio region has an elongation southward toward the dynamical center. 

%% fig. 7
\begin{figure}
  \begin{center}
    \FigureFile(80mm,76mm){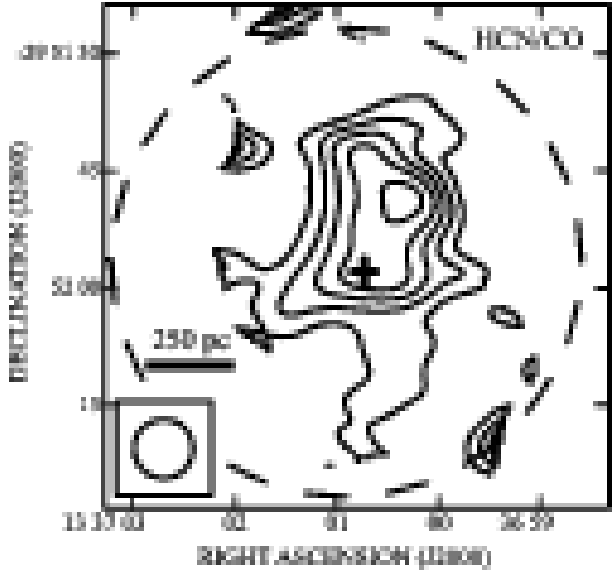}
  \end{center}
\caption{Contour map of $R_{\rm HCN/CO}$ in the central region of M~83.
The central cross indicates the dynamical center.
The contour levels are 0.02, 0.04, 0.06, 0.08, and 0.10, and the peak is 0.11.
The peak of $R_{\rm HCN/CO}$ resides northward with respect to the dynamical center.
The spatial resolution of this map is 7$^{\prime \prime}$.5 $\times$ 7$^{\prime \prime}$.5,
indicated by a circle in the lower left corner. The dashed circle represents the FOV of the CO image. 
}
\label{fig:fig7}
\end{figure}
%% fig. 7

\subsection{Comparison with optical image}

To compare our millimeter-wave images with an optical image obtained with the $HST$/WFPC2,
we made the maps of the combined CO($J=1-0$) intensity, the HCN($J=1-0$) intensity, the 95~GHz continuum,
and $R_{\rm HCN/CO}$, respectively, superposed on the HST V-band image (F547M)
obtained from $HST$/WFPC2 archival data.
These maps are shown in figure~8.
The V-band image shows the distribution of optically luminous young star clusters,
which is referred to as ``optical starburst.''
The optical starburst region is distributed on the south side of the center,
and is confined between the twin peaks in the CO($J=1-0$) emission line.
The distribution of the HCN($J=1-0$) integrated intensity is shifted northward with respect to
that of the optical starburst. Moreover, even the 95~GHz continuum, which is believed to trace
current star formation as well as H$\alpha$, is also clearly shifted northward
from the optical starburst region and the peak is close to that in the HCN($J=1-0$) line.
This means the ``optical starburst'' seen in the V-band does not trace current star-forming region.
The fact is also reported on the basis of near-IR imaging (\cite{ryder2005}, \cite{houghton2008}).

%% fig. 8
\begin{figure}
  \begin{center}
    \FigureFile(170mm,173mm){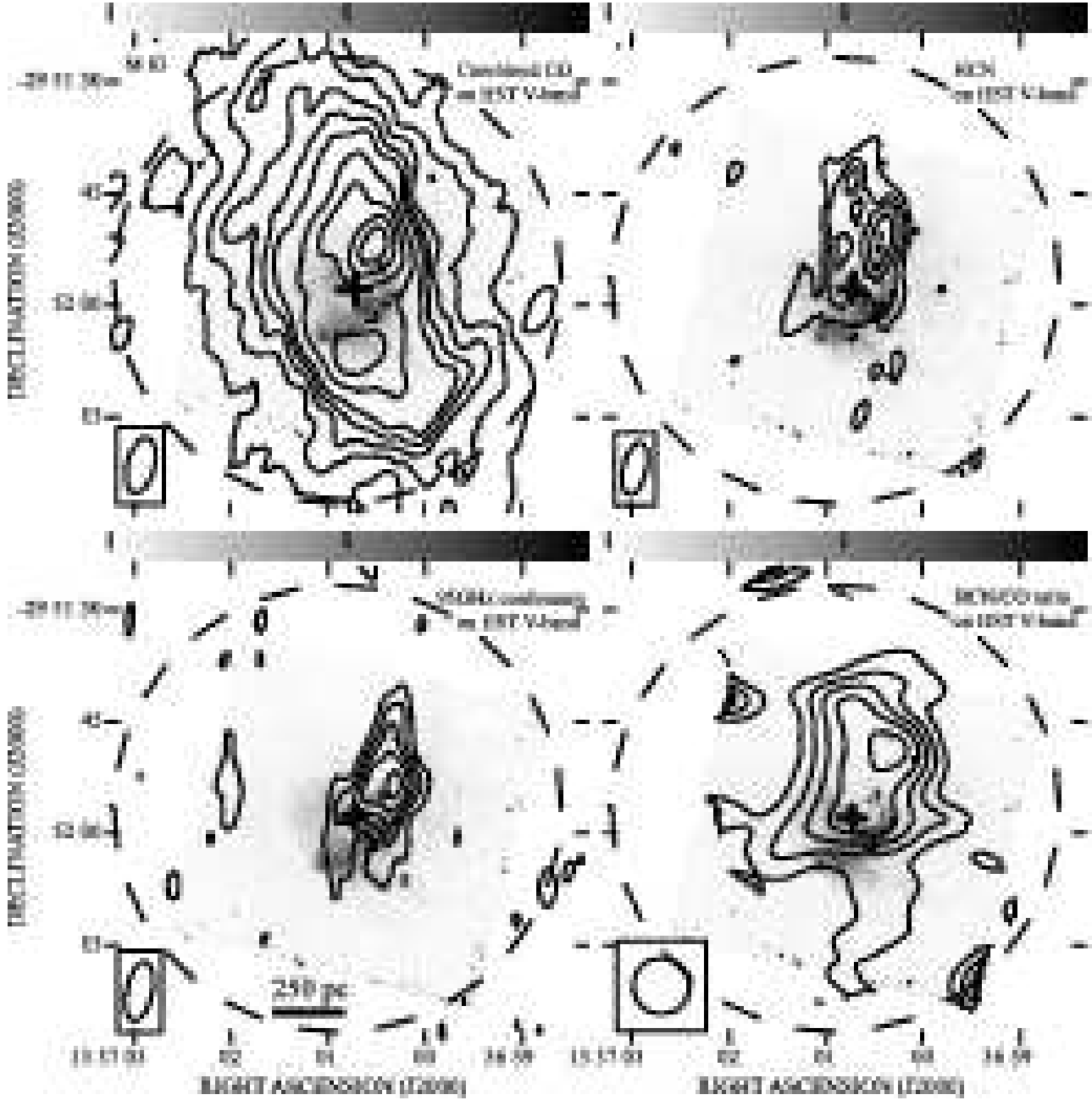}
  \end{center}
\caption{(top left) Integrated intensity map of the combined CO($J=1-0$) (contour) superposed on
the HST V-band (F547M) image (grey scale). The contour levels are the same as those in the right of figure~5.
The central cross indicates the dynamical center.
(top right) Integrated intensity map in the HCN($J=1-0$) (contour) superposed on
the HST V-band image (grey scale). The contour levels are the same as those in the bottom left of figure~2.
(bottom left) 95~GHz continuum image (contour) superposed on
the HST V-band image (grey scale). The contour levels are the same as those in the bottom right of figure~2.
(bottom right) $R_{\rm HCN/CO}$ map (contour) superposed on
the HST V-band image (grey scale). The contour levels are the same as those in figure~7.
}
\label{fig:fig8}
\end{figure}
%% fig. 8

\section{Discussion}

\subsection{Extinction in the central region of M~83}

As described in the previous section, we found that the optical starburst region of M~83 is not
spatially coincident with current star-forming region traced by the 95~GHz continuum.
This spatial inconsistency is probably due to strong extinction.
To confirm this possibility, we estimate the magnitude and distribution of extinction.

Some previous researches conclude that extinction at the center of M~83 is not small.
\citet{thatte2000} reported that V-band extinction $A_{\rm V}$ spreads from 0.5 to 9.2 mag
in the central 12$^{\prime \prime}$ ($\sim$ 250 pc) region.
\citet{harris2001} reported that extinction to the north of the center ($\sim 15^{\prime \prime}$ or 330 pc) is
stronger than that to the south.
This suggests that $A_{\rm V}$ is different from place to place and it is very large.
Therefore, we should estimate the $A_{\rm V}$ with enough accuracy. 

To estimate the H$\alpha$ extinction, $A_{\rm H \alpha}$, we use the Pa$\alpha$ emission.
We employed an H$\alpha$ image obtained with the CTIO 1.5-m telescope \citep{meurer2006}
and a Pa$\alpha$ image obtained with archival data
of the {\it HST}/NICMOS camera (P.I. M. Rieke, Proposal I.D. 7218).
The Pa$\alpha$ image is convolved to be the same resolution as that of the H$\alpha$, $1^{\prime \prime}.84$,
although the resolution of the Pa$\alpha$ image is $0.^{\prime \prime}15$.
Note that the FOV of the Pa$\alpha$ image is narrower than that of NMA observations.

We adopted a metallicity-dependent intrinsic ratio H$\alpha$/Pa$\alpha$ = 8.45
in the assumption of electron temperature $T_e$ = 10000 K for $n_e = 100$ ${\rm cm}^{-3}$ \citep{osterbrock2006}.
This means that without any extinction
\begin{eqnarray}
L_{\rm H \alpha, corr}/L_{\rm Pa \alpha, corr} = 8.45,
\end{eqnarray}
where $L_{\rm H \alpha, corr}$ means extinction-corrected H$\alpha$ luminosity, and $L_{\rm Pa \alpha, corr}$ means
that of Pa$\alpha$.
From an extinction curve we used,
\begin{eqnarray}
A_{\rm H \alpha}/A_{\rm Pa \alpha} = 6.0.
\end{eqnarray}
From combination of these two relations, we got formula to calculate $A_{\rm H \alpha}$.
\begin{eqnarray}
L_{\rm H \alpha, corr}/L_{\rm Pa \alpha, corr} &=& (10^{A_{\rm H \alpha}/2.5} L_{\rm H \alpha, obs}) / (10^{A_{\rm Pa \alpha}/2.5} L_{\rm Pa \alpha, obs}) \nonumber \\
                                               &=& (10^{A_{\rm H \alpha}/2.5} L_{\rm H \alpha, obs}) / (10^{A_{\rm H \alpha}/15} L_{\rm Pa \alpha, obs}) \nonumber \\
                                               &=& 8.45.
\end{eqnarray}
Since $L_{\rm H \alpha, obs}$ and $L_{\rm Pa \alpha, obs}$ can be calculated from each observed set of data,
$A_{\rm H \alpha}$ could be obtained.

Figure~9 shows the obtained $A_{\rm H \alpha}$ map and an extinction-corrected H$\alpha$ luminosity map
for the central 20$^{\prime \prime}$ region of M~83.
The maximum $A_{\rm H \alpha}$ is about 4 mag, and its location is almost coincident with
the nuclear peak of the HCN($J=1-0$) emission.
$A_{\rm H \alpha}$ $\sim$ 4 mag corresponds to $A_{\rm V}$ $\sim$ 5 mag.
Thus, the optical light is extinct by a factor of 100 around the HCN($J=1-0$) peak.
In addition, the peak of the extinction-corrected H$\alpha$ luminosity coincides well with
those of the $A_{\rm H \alpha}$ and the CO($J=1-0$) line, the HCN($J=1-0$) line, and the 95~GHz continuum emission.
On the other hand, extinction is almost negligible to the south of the center.
This inhomogeneous extinction shows why the optical starburst is clearly visible
to the south of the center but almost invisible in the north side.
This suggests the existence of deeply buried ongoing starburst with strong extinction,
which is already reported by several authors (e.g., \cite{ryder2005}, \cite{houghton2008}),
near the peaks in the HCN($J=1-0$) line and the 95~GHz continuum emission.

%% fig. 9
\begin{figure}
  \begin{center}
    \FigureFile(160mm,171mm){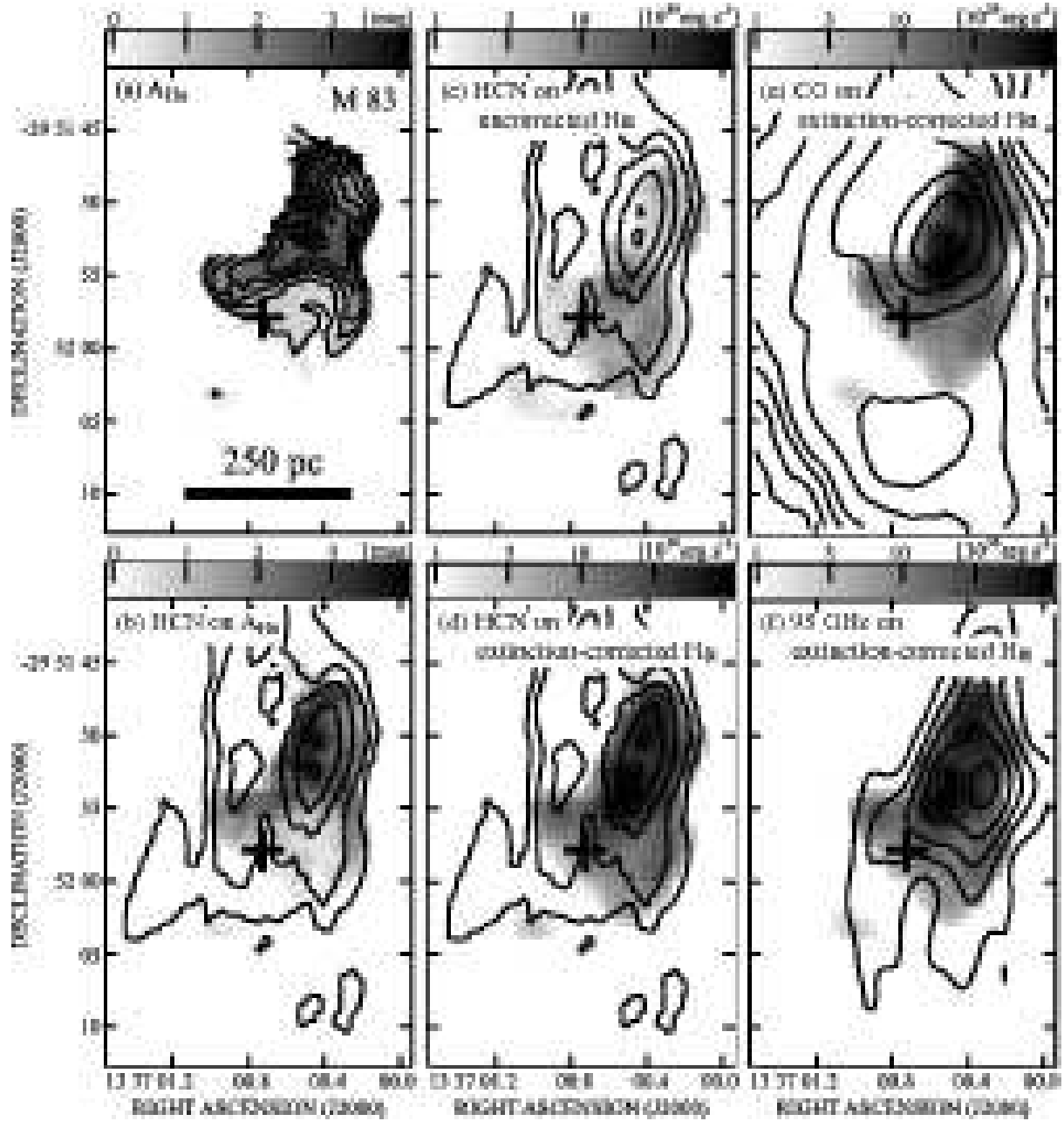}
  \end{center}
\caption{(a) A map of extinction in H$\alpha$ emission.
The contour levels are 0.5, 1.0, 1.5, 2.0, 2.5, 3.0, and 3.5 mag,
and the peak is 3.9 mag.
The central cross indicates dynamical center.
(b) An integrated intensity map in the HCN($J=1-0$) emission (contour)
superposed on an extinction map of H$\alpha$ (grey scale).
The contour levels are the same as those of figure~2.
(c) An integrated intensity map in the HCN($J=1-0$) emission (contour)
superposed on an uncorrected H$\alpha$ luminosity map (grey scale).
(d) An integrated intensity map in the HCN($J=1-0$) emission (contour)
superposed on an extinction-corrected H$\alpha$ luminosity map (grey scale) 
(e) An integrated intensity map of the combined CO($J=1-0$) emission (contour)
superposed on an extinction-corrected H$\alpha$ luminosity map (grey scale).
The contour levels are the same as those of figure~2.
(f) A map of the 95~GHz continuum emission (contour)
superposed on an extinction-corrected H$\alpha$ luminosity map (grey scale).
The contour levels are the same as those of figure~2.
}
\label{fig:fig9}
\end{figure}
%% fig. 9

\subsection{Star formation rates at the center of M~83}

In order to discuss the relationship between dense molecular gas and star formation,
we estimate the SFR in the central $22^{\prime \prime}$ ($\sim$ 500 pc) region of M~83.
We used various SFR indicators, i.e. our 95~GHz (3 mm) continuum emission, 5 GHz (6 cm) continuum emission,
infrared (IR) luminosity, and extinction-corrected H$\alpha$ luminosity.
All of them should be consistent when all corrections are applied properly.

\subsubsection{SFR derived from thermal free-free emission flux}

Continuum emission at 95~GHz is expected to be an extinction-free tracer of SFR.
This is because continuum emission in this wavelength is dominated by thermal free-free emission,
and is therefore directly converted to the Lyman photon rate \citep{condon1992}.

By assuming that the observed 95~GHz continuum is dominated by the thermal free-free emission,
a Lyman continuum rate or H$\alpha$ luminosity can be evaluated by the following formula (\cite{condon1992}, \cite{kohno2008b}),
\begin{eqnarray}
L_{{\rm H} \alpha} = 9.6 \times 10^{37} \left( \frac{D}{\rm Mpc} \right)^2 \left( \frac{T_e}{10^4\,\,{\rm K}} \right) \left( \frac{\nu}{\rm GHz} \right)^{0.1} \left( \frac{S_{\rm thermal}}{\rm mJy} \right) \,\, {\rm erg} \,\, {\rm s^{-1}}
\end{eqnarray}
where $D$ is the distance to the galaxy, $T_e$ is the electron temperature, $\nu$ is the frequency,
and $S_{\rm thermal}$ is the flux density of the thermal free-free continuum emission.
We assume that the observed 95~GHz continuum flux within the central $22^{\prime \prime}$ region of M~83,
$S_{\rm thermal}$ = 30 mJy, is dominated by the thermal free-free continuum emission.
Then, we derived $L_{{\rm H} \alpha}$(95~GHz) = 9.2 $\times$ $10^{40}$ erg s$^{-1}$.
Using this $L_{{\rm H} \alpha}$(95~GHz), we can derive an SFR by adopting the relation between ${{\rm H} \alpha}$ luminosity and SFR (\cite{kennicutt1998a}, \cite{kennicutt1998b}),
\begin{eqnarray}
{\rm SFR} = 7.9 \times 10^{-42} \left( \frac{L_{{\rm H} \alpha}}{{\rm erg} \,\, {\rm s}^{-1}} \right) M_{\odot} \,\, {\rm yr}^{-1}.
\end{eqnarray}
The resultant SFR from our 95~GHz continuum flux is 0.73 $\pm$ 0.21 $M_{\odot} \,\, {\rm yr}^{-1}$.
The error is estimated from the signal-to-noise(S/N) ratio of the 95~GHz continuum map.

\subsubsection{SFRs derived from IR luminosity and non-thermal radio continuum flux}

In order to evaluate the validity of the derived SFR from 95~GHz continuum flux,
we computed the SFR from IR luminosity and 5 GHz radio continuum flux.
According to \citet{kennicutt1998a}, the SFR derived from IR luminosity within the central 22$^{\prime \prime}$ region is 0.37 $M_{\odot} \,\, {\rm yr}^{-1}$,
which is almost half the value of the SFR based on the 95~GHz continuum.

We then derived the SFR from 5 GHz radio continuum flux.
Non-thermal radio luminosity is related to the observed radio continuum flux density as
\begin{eqnarray}
L_{\rm non-thermal} = 1.2 \times 10^{17} \left( \frac{D}{\rm Mpc} \right)^2 \left( \frac{S}{\rm mJy} \right) \left[1- \left[ 1 + 10 \left( \frac{\nu}{\rm GHz} \right)^{(0.1-\alpha)} \right]^{-1} \right] \, {\rm W} \, {\rm Hz}^{-1}
\end{eqnarray}
(\cite{kohno2008b}), where $D$ is the distance, $S$ the observed flux density at the frequency of $\nu$, and $\alpha$ the non-thermal continuum spectral index ($\sim$ 0.8).
This non-thermal radio luminosity is related to an SFR as
\begin{eqnarray}
{\rm SFR} = 8.2 \times 10^{-22} \left( \frac{\nu}{\rm GHz} \right)^{\alpha} \left( \frac{L_{\rm non-thermal}}{{\rm W} \,\, {\rm Hz}^{-1}} \right) M_{\odot} \,\, {\rm yr}^{-1}
\end{eqnarray}
(\cite{jogee2005}). From these equations and the 5 GHz radio continuum map produced by \citet{neininger1993},
we obtained the SFR of 0.73 $\pm$ 0.15 $M_{\odot} \,\, {\rm yr}^{-1}$ within the central 22$^{\prime \prime}$ region,
which is the same value as the SFR based on the 95~GHz continuum.

\subsubsection{SFR derived from extinction-corrected H$\alpha$ luminosity}

SFR can be calculated from H$\alpha$ luminosity as shown in equation (5).
However, the H$\alpha$ emission often suffers from extinction by interstellar dust.
In fact, there is up to 4 mag of H$\alpha$ extinction in the central region of M~83 as described in the previous subsection.

Therefore, there is no doubt that appropriate correction of extinction is indispensable.
Here we must verify what data should be used to correct extinction.
Pa$\alpha$ seems to be very useful, but cannot cover the entire FOV of CO($J=1-0$) and HCN($J=1-0$) image.

Recently, the Spitzer/MIPS 24 $\mu$m image has begun to be employed
for calibration of SFR (e.g. M~51; \cite{calzetti2007}).
An archival MIPS 24 $\mu$m image (P00059, George, Rieke, Starburst activity in nearby galaxies)
covers the entire disk of M~83, and the spatial resolution of the image is about 5$^{\prime \prime}$.7.
The formula to calibrate H$\alpha$ luminosity using 24 $\mu$m image is as follows \citep{calzetti2007}.
\begin{eqnarray}
L_{\rm H \alpha, corr} = L_{\rm H \alpha, obs} + (0.031 \pm 0.006) L_{\rm 24 \mu m} \,\,\, {\rm erg} \,\, {\rm s^{-1}}
\end{eqnarray}
where $L_{\rm H \alpha, obs}$ means observed H$\alpha$ luminosity, and
$L_{\rm 24 \mu m}$ means that of 24 $\mu$m.
At the center of M~83, H$\alpha$ extinction derived from MIPS 24 $\mu$m image is about 3 mag
at a resolution of 5$^{\prime \prime}$.7.
Considering the difference in spatial resolution, the derived H$\alpha$ extinction from MIPS 24 $\mu$m image
seems to be consistent with that from the Pa$\alpha$ image.
Using equation (5), the resultant SFR from extinction-corrected H$\alpha$ is 0.24 $\pm$ 0.05 $M_{\odot} \,\, {\rm yr}^{-1}$
within the central 22$^{\prime \prime}$ region. This value is close to that from IR luminosity,
but 3 times smaller than that from 95~GHz and 5 GHz continuum flux.

The SFRs derived from various indicators are summarized in table~6.
It is unclear what causes such a significant discrepancy among these SFRs.

\begin{table}
\begin{center}
Table~6.\hspace{4pt}Star formation rates from various indicators within the central 22$^{\prime \prime}$ region.\\[1mm]
\begin{tabular}{ccc}
\hline \hline \\[-5mm]
Indicator & SFR & Reference of emission data\\
\hline \\[-5mm]
95~GHz (3 mm) continuum                   & 0.73 $\pm$ 0.21 & This work \\
5 GHz  (6 cm) continuum                   & 0.73 $\pm$ 0.15 & (1)       \\
IR luminosity                    & 0.37            & (2)       \\
extinction-corrected H$\alpha$   & 0.24 $\pm$ 0.05 & (3)       \\
\hline \\[-2mm]
\end{tabular}\\
{\footnotesize
Reference\,---. (1) \citet{neininger1993}, (2) \citet{kennicutt1998a}, (3) \citet{meurer2006}
}
\end{center}
\end{table}

\subsection{Comparison between $R_{\rm HCN/CO}$ and SFE}

\subsubsection{Derivation of SFE}

In order to compare the SFE in the central region directly with our $R_{\rm HCN/CO}$ data,
we need reliable SFR data with an adequate sensitivity and an adequate spatial resolution higher than $7^{\prime \prime}.5$.
For the IR luminosity data and the 5 GHz continuum data, their spatial resolutions ($\geq$ 10$^{\prime \prime}$) are inadequate.
In addition, our 95~GHz continuum data is unfavorable for comparing $R_{\rm HCN/CO}$
since the area where 95~GHz continuum emission is detected in adequate S/N ratio (more than 5 $\sigma$)
is narrower than that of $R_{\rm HCN/CO}$.
Then, we use the SFR data based on the extinction-corrected H$\alpha$ luminosity using MIPS 24 $\mu$m data
in order to calculate the SFE in the central region.

The SFE is calculated as follows,
\begin{eqnarray}
{\rm SFE} = \left( \frac{\Sigma_{\rm SFR}}{M_{\odot} \,\, {\rm yr^{-1} \,\, pc^{-2}}} \right) \left( \frac{\Sigma_{\rm H_2}}{M_{\odot} \,\, {\rm pc^{-2}}} \right)^{-1} \,\,\, {\rm yr^{-1}}.
\end{eqnarray}
$\Sigma_{\rm SFR}$ is the surface mass density of SFR, and $\Sigma_{\rm H_2}$ is that of molecular gas.
$\Sigma_{\rm H_2}$ is calculated as follows,
\begin{eqnarray}
{\Sigma_{\rm H_2}} = 2.89 \,\, {\rm cos} i \,\, \left( \frac{I_{\rm CO(J=1-0)}}{\rm K \,\, km \,\, s^{-1}} \right) \,\,\, M_{\odot} \,\, {\rm pc^{-2}}.
\end{eqnarray}

Here, the $N_{\rm H_2}$/$I_{\rm CO}$ conversion factor ($X_{\rm CO}$) is assumed to be $1.8 \times 10^{20}$ cm$^{-2}$$({\rm K \,\, km \,\, s^{-1}})^{-1}$ \citep{dame2001}.
$I_{\rm CO(J=1-0)}$ is derived from the combined CO($J=1-0$) intensity map.
Figure~10 shows the calibrated $\Sigma_{\rm SFR}$ map and the SFE map.
Both maps are convolved to the resolution of $7^{\prime \prime}.5 \times 7^{\prime \prime}.5$ (160 pc $\times$ 160 pc)
to match $R_{\rm HCN/CO}$ data. The peak value of the SFR is $\sim 4 \times 10^{-6} \,\, M_{\odot}\,{\rm yr^{-1}}$.
The peak value of SFE (i.e. the site where most active star formation is supposed to occur) is
$\sim 5 \times 10^{-9} \,\, {\rm yr^{-1}}$ and it is located north of the center.
The highest SFE region is spatially coincident well with the highest SFR region.

%% fig. 10
\begin{figure}
  \begin{center}
    \FigureFile(170mm,89mm){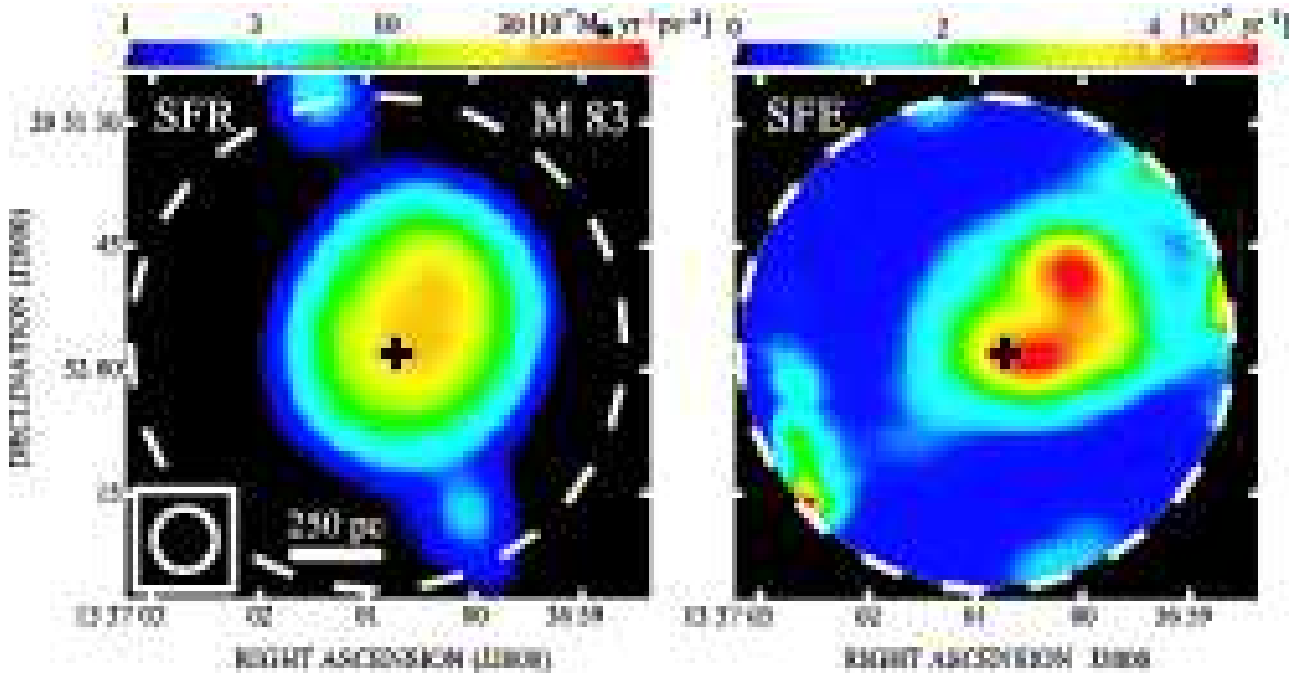}
  \end{center}
\caption{(left) A map of the SFR in the central region of M~83 estimated from calibrated H$\alpha$ luminosity. 
The central cross indicates the dynamical center.
The peak of the SFR is $4 \times 10^{-6}$ $M_{\odot} \,\, {\rm yr^{-1} \,\, pc^{-2}}$.
The spatial resolution of this map is 7$^{\prime \prime}$.5 $\times$ 7$^{\prime \prime}$.5,
indicated by a circle in the left corner.
(right) A map of the SFE in the central region of M~83.
The peak of the SFE is $5 \times 10^{-9}$ ${\rm yr^{-1}}$, and it coincides well spatially with that of SFR.
}
\label{fig:fig10}
\end{figure}
%% fig. 10

\subsubsection{$R_{\rm HCN/CO}$ vs.\ SFE in M~83: correlation in a GMA scale}

Here, we compare $R_{\rm HCN/CO}$ with the SFE within the central 1 kpc region of M~83.
Figure~11 shows the map of $R_{\rm HCN/CO}$ superposed on that of SFE.
The spatial correlation between these two maps seems to be roughly good,
but $R_{\rm HCN/CO}$ map seems to trail the skirt toward the northeast.
This skirt-like structure of $R_{\rm HCN/CO}$ is not seen in the SFE map,
and is coincident with the dust lane seen in the VLT V-band image \citep{comeron2001}.

%% fig. 11
\begin{figure}
  \begin{center}
    \FigureFile(80mm,81mm){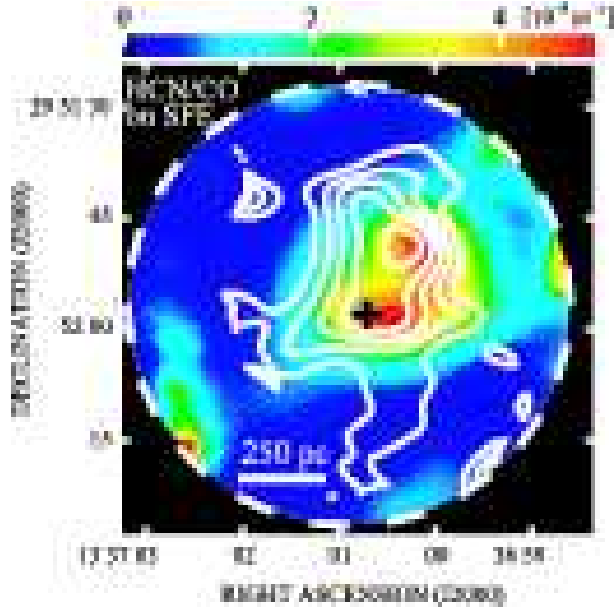}
  \end{center}
\caption{A map of $R_{\rm HCN/CO}$ (white contour) superposed on a map of the SFE in the central region of M~83 (color).
The central cross indicates the dynamical center.
The spatial correlation between these two maps seems to be roughly good.
The contour levels are the same as those of figure~7.
}
\label{fig:fig11}
\end{figure}
%% fig. 11

In addition, we examine the correlation between $R_{\rm HCN/CO}$ and the SFE.
For the region where $R_{\rm HCN/CO}$ exceeds 0.02, the values of $R_{\rm HCN/CO}$ and the SFE are obtained
for each separation of $3^{\prime \prime}.75$, which corresponds to half of the spatial resolution of each map.
Figure~12 shows a plot of $R_{\rm HCN/CO}$ vs.\ the SFE in each region.
The correlation between $R_{\rm HCN/CO}$ and the SFE is clearly seen.

These $R_{\rm HCN/CO}$ correspond to the dense gas fraction in the central region of M~83
on a 160 pc (corresponding to GMA) scale.
The dense gas fraction would be translated as the number of star-forming dense cores per unit gas mass.
Then, the correlation derived between $R_{\rm HCN/CO}$ means that 
an outbreak of extensive star formation (high-SFE star formation) such as the nuclear starburst
requires the generation of a large number of star-forming dense cores within GMAs
(e.g., \cite{solomon1992}, \cite{kohno2002}, \cite{shibatsuka2003}, \cite{gao2004a}).

\subsubsection{$R_{\rm HCN/CO}$ vs.\ SFE: comparison with a global scale correlation}

We compare our results with that shown by \citet{gao2004a}.
We converted the SFR in M~83, which is estimated from extinction-corrected H$\alpha$ luminosity,
to the total IR (8 to 1000 $\mu$m) luminosity using the following formula \citep{kennicutt1998a},
\begin{eqnarray}
L_{\rm IR} =  2.2 \times 10^{43} \left( \frac{\rm SFR}{M_{\odot} \,\,{\rm yr}^{-1}} \right) {\rm erg} \,\, {\rm s}^{-1}.
\end{eqnarray}
Then, we adapted the vertical axis of our $R_{\rm HCN/CO}$ vs.\ SFE plot to that of figure~5a in \citet{gao2004a}.
Figure~13 shows the composite of the $R_{\rm HCN/CO}$ vs.\ SFE plot for the center of M~83
and for ULIRGs, LIRGs, and normal spirals \citep{gao2004a}.
The correlation between $R_{\rm HCN/CO}$ and the SFE in the central region of M~83
almost seems to coincide with that of \citet{gao2004a} sample.
This suggests that the correlation between $R_{\rm HCN/CO}$ and SFE on a GMA ($\sim$ 160 pc) scale
found in the nuclear starburst region of M~83 is the origin of the global correlation
on a galactic (a few kpc) scale shown by \citet{gao2004a}.
In other words, $R_{\rm HCN/CO}$ (dense gas fraction) and SFE on a galactic scale
are averages of those parameters on a GMA scale.
Low-$R_{\rm HCN/CO}$ (less dense) GMAs would be dominant in a low-SFE galaxy,
whereas high-$R_{\rm HCN/CO}$ (dense) GMAs are possibly dominant in a high-SFE galaxy.
This is consistent with the prediction of a three-dimensional, high-resolution hydrodynamic simulation \citep{wada2007}.
They showed the SFR and SFE are sensitive to increasing average density of molecular gas.
The average gas density just corresponds to $R_{\rm HCN/CO}$.

%% fig. 12
\begin{figure}
  \begin{center}
    \FigureFile(80mm,70mm){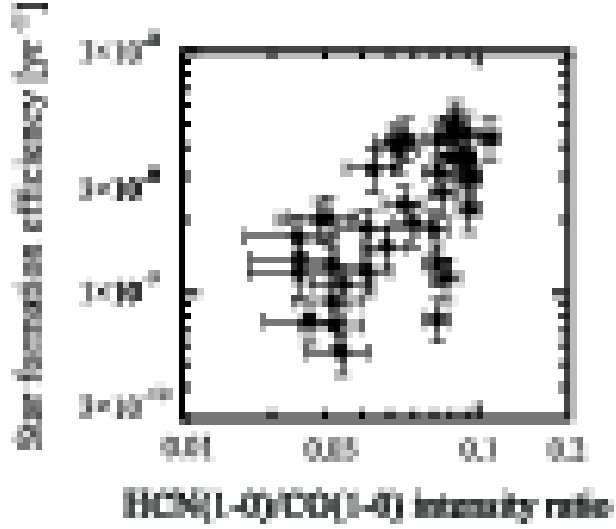}
  \end{center}
\caption{A plot of $R_{\rm HCN/CO}$ vs.\ SFE in the central region of M~83. 
The correlation between $R_{\rm HCN/CO}$ and the SFE is clearly seen
in $7^{\prime \prime}.5 \times 7^{\prime \prime}.5$ (160 pc $\times$ 160 pc) scale.
}
\label{fig:fig12}
\end{figure}
%% fig. 12

%% fig. 13
\begin{figure}
  \begin{center}
    \FigureFile(80mm,80mm){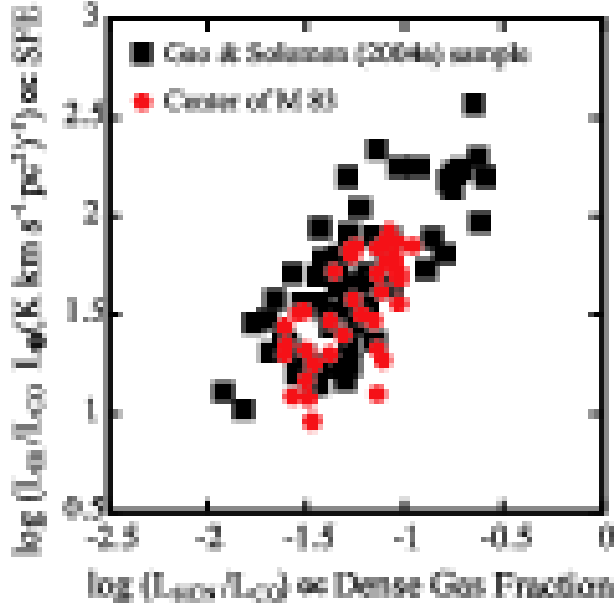}
  \end{center}
\caption{A composite of the $R_{\rm HCN/CO}$ vs.\ SFE plots for the central region of M~83 (this work)
and for ULIRGs, LIRGs, and normal spirals (\cite{gao2004a}).
The distances of the \citet{gao2004a} sample are in the range of 2.5 Mpc to 266 Mpc.
Black squares indicate the \citet{gao2004a} sample, and red circles indicate the center of M~83.
The correlation between $R_{\rm HCN/CO}$ and the SFE in the central region of M~83 almost
seems to coincides with that of the \citet{gao2004a} sample.
}
\label{fig:fig13}
\end{figure}
%% fig. 13

\section{Summary}

We have performed aperture synthesis high-resolution ($\sim 7^{\prime \prime} \times 3^{\prime \prime}$) observations in
the CO($J=1-0$) line, the HCN($J=1-0$) line, and the 95~GHz continuum emission toward the central region ($\sim$ 1.5 kpc) of
the nearby barred spiral galaxy M~83 with the NMA.
A summary of this work is as follows.

\begin{enumerate}
\item The size of the CO($J=1-0$) map is 2$^{\prime}$ $\times$ 1$^{\prime}$ (3 pointings mosaic observations).
The synthesized beam size and the resultant rms noise level of the intensity map are
7$^{\prime \prime}$ $\times$ 3$^{\prime \prime}$ and 2.7 Jy\,beam$^{-1}$\,km\,s$^{-1}$, respectively.
Our high resolution CO($J=1-0$) mosaic map with the highest sensitivity and the highest spatial resolution
to date depicts the presence of molecular ridges along the leading sides of the stellar bar and nuclear twin peak structure.
In addition, we combined the NMA CO($J=1-0$) data with the NRO 45-m CO($J=1-0$) data.
The combined CO($J=1-0$) map first reveals the high-resolution distribution of molecular gas
containing diffuse components in the central region of M~83.

\item The size of the HCN($J=1-0$) map and the 95~GHz continuum map is 77 $^{\prime \prime}$.
The synthesized beam size is 7$^{\prime \prime}$ $\times$ 3$^{\prime \prime}$ for the HCN($J=1-0$) map
and 8$^{\prime \prime}$ $\times$ 3$^{\prime \prime}$ for the 95~GHz continuum map, respectively.
The resultant noise level is 0.52 Jy\,beam$^{-1}$\,km\,s$^{-1}$ for the HCN($J=1-0$) intensity map
and 0.85 mJy\,beam$^{-1}$ for 95~GHz continuum map, respectively.
We found the distribution of the HCN($J=1-0$) line emission which traces dense molecular gas
shows nuclear single peak structure, and coincides well
with that of the 95~GHz continuum emission which traces massive starburst.
However, the peaks of the HCN($J=1-0$) line and the 95~GHz continuum emission are not associated with
the optical starburst traced by the HST V-band image.

\item Using the H$\alpha$/Pa$\alpha$ ratio, an extinction map of the center of M~83 is obtained.
The highest extinction is $A_{\rm H \alpha} \sim 4$ mag ($A_{\rm V} \sim 5$ mag),
and which is spatially coincides with the peak of extinction-corrected H$\alpha$ luminosity
and that of the HCN($J=1-0$) line emission.
This suggests the existence of deeply buried ongoing starburst due to strong extinction
near the peaks of the HCN($J=1-0$) line and the 95~GHz continuum emission.

\item We found that $R_{\rm HCN/CO}$ correlates well with the extinction-corrected SFE using the MIPS 24 $\mu$m data
in the central region of M~83 at a resolution of $7^{\prime \prime}.5$ ($\sim 160$ pc).
That is, the SFE is controlled by dense gas fraction traced by $R_{\rm HCN/CO}$ on a GMA scale.
In addition, the correlation between $R_{\rm HCN/CO}$ and the SFE in the central region of M~83 seems to be
almost coincident with that of \citet{gao2004a} sample.
This suggests that the correlation between $R_{\rm HCN/CO}$ and the SFE on a GMA ($\sim$ 160 pc) scale
found in the nuclear starburst region of M~83 is the origin of the global correlation
on a few kpc scale shown by \citet{gao2004a}.
\end{enumerate}

\vspace{0.5cm}

We would like to acknowledge the referee for his invaluable comments.
We are deeply indebted to the NRO staff for the operation of the telescopes
and their continuous efforts to improve the performance of the instruments.
We are grateful to F. Comer{\'o}n for sending us the V-band image of M~83 obtained with VLT,
M. Fukuhara for providing us his CO($J=1-0$) image obtained with the NRO 45-m telescope,
and A. Hirota for his HCN($J=1-0$) data obtained with the NRO 45-m.
K. M. was financially supported by a Grant-in-Aid for JSPS Fellows.
This study was partly supported by the MEXT Grant-in-Aid for Scientific Research on Priority Areas No.\ 15071202.
This research has made use of the NASA/IPAC Extragalactic Database (NED) which is operated by the Jet Propulsion Laboratory,
California Institute of Technology, under contract with the National Aeronautics and Space Administration
This work is based on observations made with the NASA/ESA Hubble Space Telescope,
obtained from the data archive at the Space Telescope Science Institute.
STScI is operated by the Association of Universities for Research in Astronomy, Inc. under NASA contract NAS 5-26555.
This work is based on observations made with the Spitzer Space Telescope,
which is operated by the Jet Propulsion Laboratory, California Institute of Technology under a contract with NASA.

%%%%%%%%%%%%%%%%%%%%%%%%%%%%%%%%%%%%%%%

%%%
% See the manual for the detail.
%%%

\end{document}